\documentclass[fleqn,11pt]{wlscirep}
\usepackage[utf8]{inputenc}
\usepackage[T1]{fontenc}
\usepackage[utopia]{mathdesign}
\usepackage[OMLmathrm,OMLmathbf]{isomath}

\newcommand{\beginsupplement}{%
        \setcounter{table}{0}
        \renewcommand{\thetable}{S\arabic{table}}%
        \setcounter{figure}{0}
        \renewcommand{\thefigure}{S\arabic{figure}}%
     }

\usepackage[numbers]{natbib}
\usepackage{setspace}
\def\LAOSTO{LaAlO$_3$/SrTiO$_3$ }

\def\LAO{LaAlO$_3$ }
\def\vsg{$V_{\mathrm{sg}}$ }

\title{One-dimensional Kronig-Penney superlattices at the LaAlO$_3$/SrTiO$_3$ interface}

\author[1,2]{Megan Briggeman}
\author[3]{Hyungwoo Lee}
\author[3]{Jung-Woo Lee}
\author[3]{Kitae Eom}
\author[2,4]{Fran\c{c}ois Damanet}
\author[2,4]{Elliott Mansfield}
\author[1,2]{Jianan Li}
\author[1,2]{Mengchen Huang}
\author[2,4]{Andrew J.~Daley}
\author[3]{Chang-Beom Eom}
\author[1,2]{Patrick Irvin}
\author[1,2,*]{Jeremy Levy}
\affil[1]{Department of Physics and Astronomy, University of Pittsburgh, Pittsburgh, PA 15260, USA}
\affil[2]{Pittsburgh Quantum Institute, Pittsburgh, PA 15260, USA}
\affil[3]{Department of Materials Science and Engineering, University of Wisconsin-Madison, Madison, WI 53706, USA}
\affil[4]{Department of Physics and SUPA, University of Strathclyde, Glasgow G4 0NG,
United Kingdom}

\affil[*]{corresponding author jlevy@pitt.edu}

\begin{abstract}

The paradigm of electrons interacting with a periodic lattice potential is central to solid-state physics \citep{Bloch1929}. 
Semiconductor heterostructures \citep{Byrnes2008,Singha2011,Slot2017} and ultracold neutral atomic lattices \citep{Jaksch2005, Kinoshita2004, Lebrat2018} capture many of the essential properties of 1D electronic systems.
However, fully one-dimensional superlattices are highly challenging to fabricate in the solid state due to the inherently small length scales involved.
Conductive atomic-force microscope (c-AFM) lithography has recently been demonstrated to create ballistic few-mode electron waveguides with highly quantized conductance and strongly attractive electron-electron interactions \citep{Annadi2018}.
Here we show that artificial Kronig-Penney-like superlattice potentials can be imposed on such waveguides, introducing a new superlattice spacing that can be made comparable to the mean separation between electrons.
The imposed superlattice potential ``fractures'' the electronic subbands into a manifold of new subbands with magnetically-tunable fractional conductance (in units of $e^2/h$).  
The lowest  $G=2e^2/h$ plateau, associated with ballistic transport of spin-singlet electron pairs \citep{Annadi2018}, is stable against de-pairing up to the highest magnetic fields explored ($|B|=$16 T).  A 1D model of the system suggests that an engineered spin-orbit interaction in the superlattice contributes to the enhanced pairing observed in the devices.
These findings represent an important advance in the ability to design new families of quantum materials with emergent properties, and mark a milestone in the development of a solid-state 1D quantum simulation platform. 
\end{abstract}

\begin{document}

\flushbottom
\maketitle

\thispagestyle{empty}

\section*{Introduction}

Quantum theory provides a unified framework for understanding the fundamental properties of matter.  However, there are many quantum systems whose behavior is not well understood because the relevant equations are are not able to be solved using known approaches.
The idea of ``quantum simulation'', first articulated by Feynman \citep{Feynman1982}, aims to exploit the quantum-mechanical properties of materials to compute the properties of interest and gain insight into the quantum nature of matter.
There are two main ``flavors'' of quantum simulation: one based upon the known efficiency of circuit-based quantum computers to solve the Schr\"odinger equation, and the other based on microscopic control over quantum systems to emulate a given Hamiltonian.  The former approach is limited by the capabilities of present-day quantum computers.  
The latter approach has shown great promise using a variety of methods including ultracold atoms 
\citep{Jaksch2005, Mazurenko2017, Choi2016}, spin systems from ion trap arrays 
\citep{Porras2004,Bohnet2016,Brydges2019}, superconducting Josephson junction arrays \citep{Houck2012,Chiesa2015,Fitzpatrick2017} photonic systems \citep{Politi2009,Peruzzo2010,Matthews2013}, and various solid-state approaches \citep{Byrnes2008,Singha2011,Slot2017,Drost2017,Kempkes2019}.  Platforms capable of quantum simulation of Hubbard models would be of enormous value in condensed matter physics and beyond.  

Complex oxides offer new opportunities to create a platform for quantum simulation in a solid-state environment.  Their complexity gives access to important quantum phases of matter, such as superconductivity, where the model Hamiltonians (e.g., 2D Hubbard model) are challenging to understand theoretically, while their nanoscale reconfigurability makes it possible to engineer new forms of quantum matter with extreme nanoscale precision \citep{Cen2008,Cen2009}. 

Here we present experiments which constitute a first step towards developing a solid-state quantum simulation platform based on a reconfigurable complex-oxide material system.  Using conductive atomic force microscope (c-AFM) lithography, we create Kronig-Penney-like \citep{Kronig1931} 1D superlattice structures by spatially modulating the potential of a 1D electron waveguide device at the \LAOSTO interface. 
Two main effects are found.
The superlattice modulation introduces new fractional conductance features that are believed to be the combined result of enhanced electron-electron interactions and the new periodic structure.  
The potential modulation also  significantly enhances the stability of spin-singlet pair transport.
The unique combination of strongly attractive electron-electron interactions, combined with the engineered properties demonstrated here, hold promise for the development of new families of quantum materials with programmable characteristics.

\begin{figure}
    \centering
    \includegraphics[width=\columnwidth]{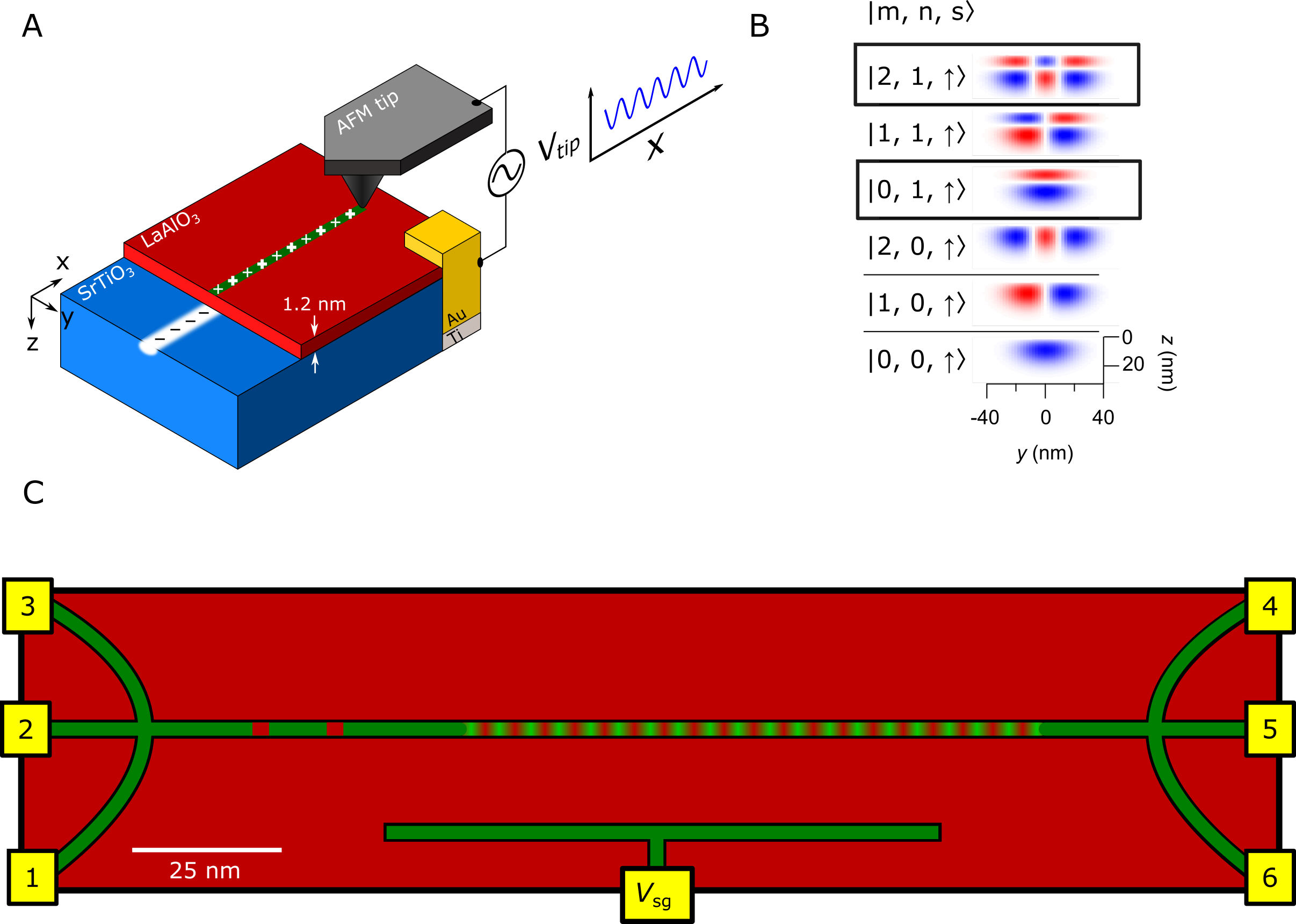}
    \caption{\textbf{Schematic of c-AFM writing and 1D superlattice device.} \textbf{(A)} C-AFM writing schematic.  A positive bias on the AFM tip protonates the \LAO surface, locally creating a conducting channel at the \LAOSTO interface.  \textbf{(B)} Chart showing different representative wavefunctions calculated using a single particle model for electron waveguide devices \citep{Annadi2018}.  The imposed vertical superlattice structure may cause mixing of vertical modes of an electron waveguide device, possibly mixing the ground state with modes highlighted in black.  \textbf{(C)} 1D vertical superlattice device schematic.  The superlattice is created by first writing the main channel with a positive tip voltage.  The same path is then traced while applying a sinusoidal tip voltage to periodically modulate the confining potential of the device.  The superlattice is created in series with two highly transparent tunnel barriers.}
    \label{fig:fig1}
\end{figure}

C-AFM lithography has been used to create a variety of devices at the \LAOSTO interface \citep{Pai2018a}.  A conductive AFM tip, moving in contact with the \LAO surface and positively-biased with respect to the \LAOSTO interface, locally creates (``writes'') conducting regions at the interface  (Figure \ref{fig:fig1}(A)), while a negatively biased tip locally restores (``erases'') the interface to an insulating state.  The mechanism behind the writing (erasing) is the local (de)protonation of the \LAO surface \citep{Bi2010,Brown2016}.  The protons on the surface create a confining potential which defines the conducting regions at the interface.  This technique achieves nanoscale control, with precision as high as 2 nm, over the conductivity of the \LAOSTO interface and most of its properties.  

The work described here concerns \LAOSTO electron waveguide devices \citep{Annadi2018} that have been perturbed by a spatially periodic potential, similar to the one first envisioned by Kronig and Penney \cite{Kronig1931}.  Unperturbed waveguides exhibit highly quantized ballistic transport with conductance steps at or near integer values of the conductance quantum $e^2/h$.  
The subband structure of \LAOSTO electron waveguides can be described by a waveguide model which takes into account vertical, lateral, and spin degrees of freedom \citep{Annadi2018}.  Representative orbitals for electron waveguides, subject to parabolic lateral confinement and half-parabolic vertical confinement, are shown in Figure \ref{fig:fig1}(B), where $|m,n,s\rangle$ is a state specified by quantum numbers $m$, $n$, and $s$ that correspond to the number of lateral ($m$) and vertical ($n$) nodes of the wavefunction, and the spin $s$.  The complete set of states form a basis for describing extended states along the waveguide direction $x$.  The periodic modulation of the waveguide may couple different vertical modes (for example those highlighed in black in Figure \ref{fig:fig1}) with the ground state $|0,0,\uparrow \rangle$. 
Due to attractive electron-electron interactions, subband energy minima can ``lock" together to form electron pairs \citep{Annadi2018} that also propagate ballistically.  
Pairing in electron waveguides arises from the same electron-electron interactions that give rise to superconductivity \citep{Cheng2015}.  
In some cases, more exotic locking of subbands can occur, e.g., the Pascal conductance plateaus which indicate the binding of $n\geq 2$ electron states \citep{Briggeman2019}.  The presence of strong, tunable electron-electron interactions makes these electron waveguide devices an interesting starting point for engineering 1D quantum systems.  

\section*{Results}

\begin{figure}
    \centering
    \includegraphics[width=0.8\columnwidth]{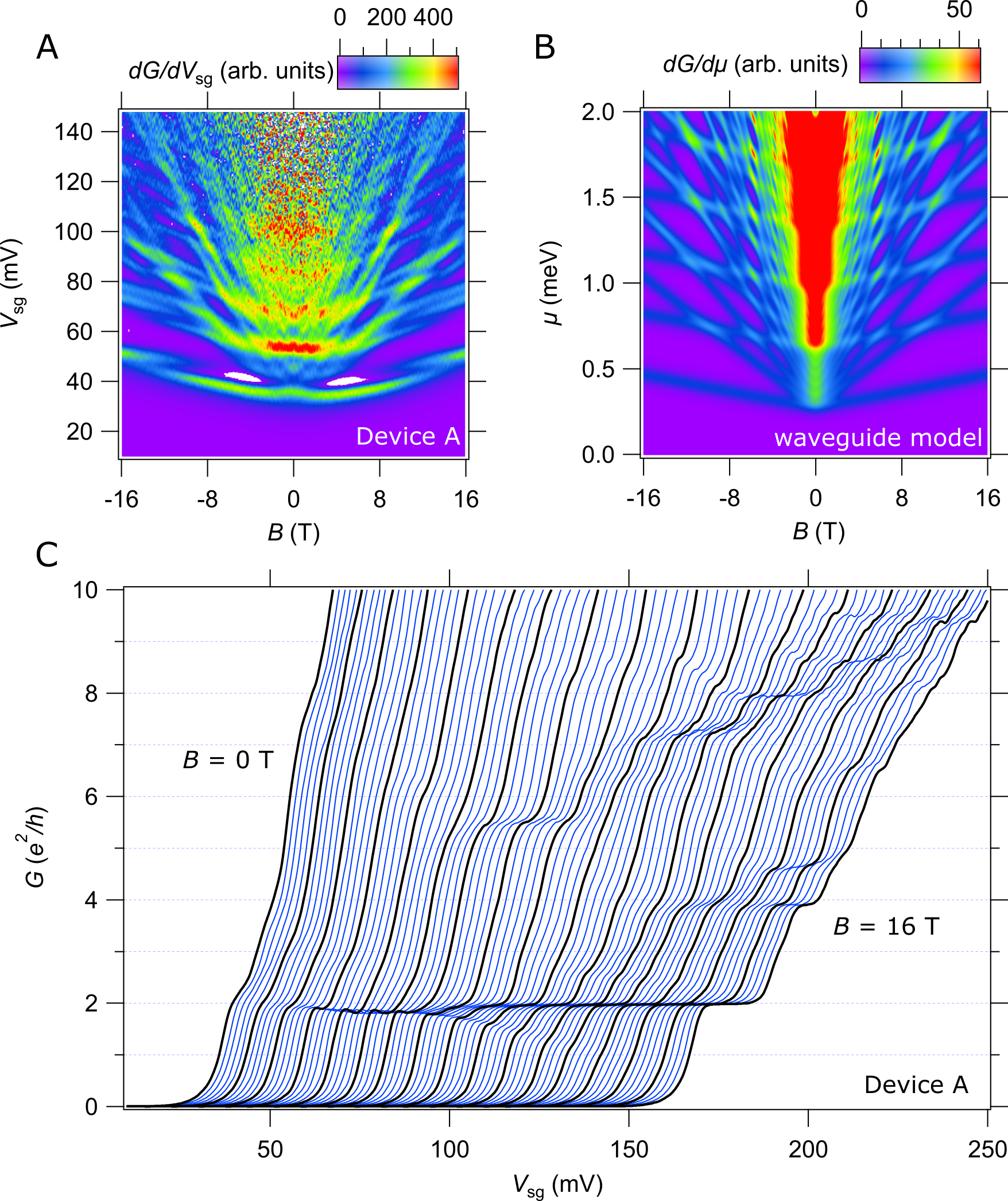}
    \caption{\textbf{Magnetotransport characteristics of a 1D vertical superlattice.}  \textbf{(A)} Transconductance $dG/dV_{\mathrm{sg}}$ as a function of magnetic field $B$ and side gate voltage $V_{\mathrm{sg}}$ for vertical superlattice Device A.  Purple regions indicate zero transconductance, or conductance plateaus.  Bright regions indicate increasing conductance when new 1D subbands become available. Negative transconductance is indicated in white, mainly in two lobes above the $2~e^2/h$ plateau around 5 T.
    \textbf{(B)} Transconductance spectra for non-interacting single-particle electron waveguide model.  
    \textbf{(C)} Conductance $G$ vs side gate voltage $V_{\mathrm{sg}}$ at magnetic fields from $B=$ 0 T to 16 T for Device A.  Curves are offset by $\Delta V_{\mathrm{sg}}\sim$ 7.5 mV/T for clarity.  Curves at 1 T intervals are highlighted in black.}
    \label{fig:fig2}
\end{figure}

1D superlattice devices are created by first writing a conductive nanowire with a constant positive voltage applied to the AFM tip ($V_{\mathrm{tip}}\sim10~\mathrm{V}$).  The same path is re-traced along the same direction while applying a sinusoidally varying tip voltage $V_{\mathrm{tip}}(x)=V_0+V_k \sin(kx)$, to produce a spatially periodic potential modulation.  A short unpatterned waveguide is written in series next to the superlattice, which helps to control the chemical potential in the device structure \citep{Annadi2018}.
Four-terminal magnetotransport measurements are carried out in a dilution refrigerator at or near its base temperature $T = 25$ mK. 
Figure \ref{fig:fig2}A shows the transconductance $dG/dV_{\textrm{sg}}$ as a function of out-of-plane magnetic field $B$ and side-gate voltage $V_{\mathrm{sg}}$ for Device A.  The transconductance map provides a visual indication of the subband structure.  Purple regions, where the transconductance is nearly zero, represent conductance plateaus.  Bright colored (red/yellow/green/blue) regions signify increases in conductance that generally correspond to the emergence of new subbands.  White regions indicate negative transconductance, resulting from an overshoot in conductance.  The transconductance is generally found to be highly symmetric with respect to the applied magnetic field.  
By comparison, Figure \ref{fig:fig2}B shows a calculated transconductance map for a single-particle model of a straight, unmodulated electron waveguide.  The non-interacting waveguide model includes the geometry of a typical electron waveguide device as well as vertical, lateral, and spin degrees of freedom, and is described in more detail elsewhere \citep{Annadi2018}.
The experimental data for the superlattice shows an overall resemblance to the waveguide model, except that the subbands are ``fractured'' into a manifold of new states with fractional conductances.
Figure \ref{fig:fig2}C shows a series of conductance curves versus \vsg for a sequence of out-of-plane magnetic fields $B$, ranging between 0 T (leftmost) to 16 T (rightmost).  Curves are offset by $\mathrm{\Delta} V_{sg}\sim$ 7.5 mV/T for clarity and curves at 1 T intervals are highlighted in black.  At low magnetic fields ($|B|\approx 2$ T), a plateau at around $1.8~e^2/h$ develops before bifurcating into two distinct plateaus, one of which decreases in value, while the other increases towards a nearly quantized value of $1.99~e^2/h$.  The onset of the two plateaus can be seen clearly in the transconductance (Figure \ref{fig:fig2}A) as a minigap that appears in the lowest subband.

\begin{figure}
    \centering
    \includegraphics[width=\columnwidth]{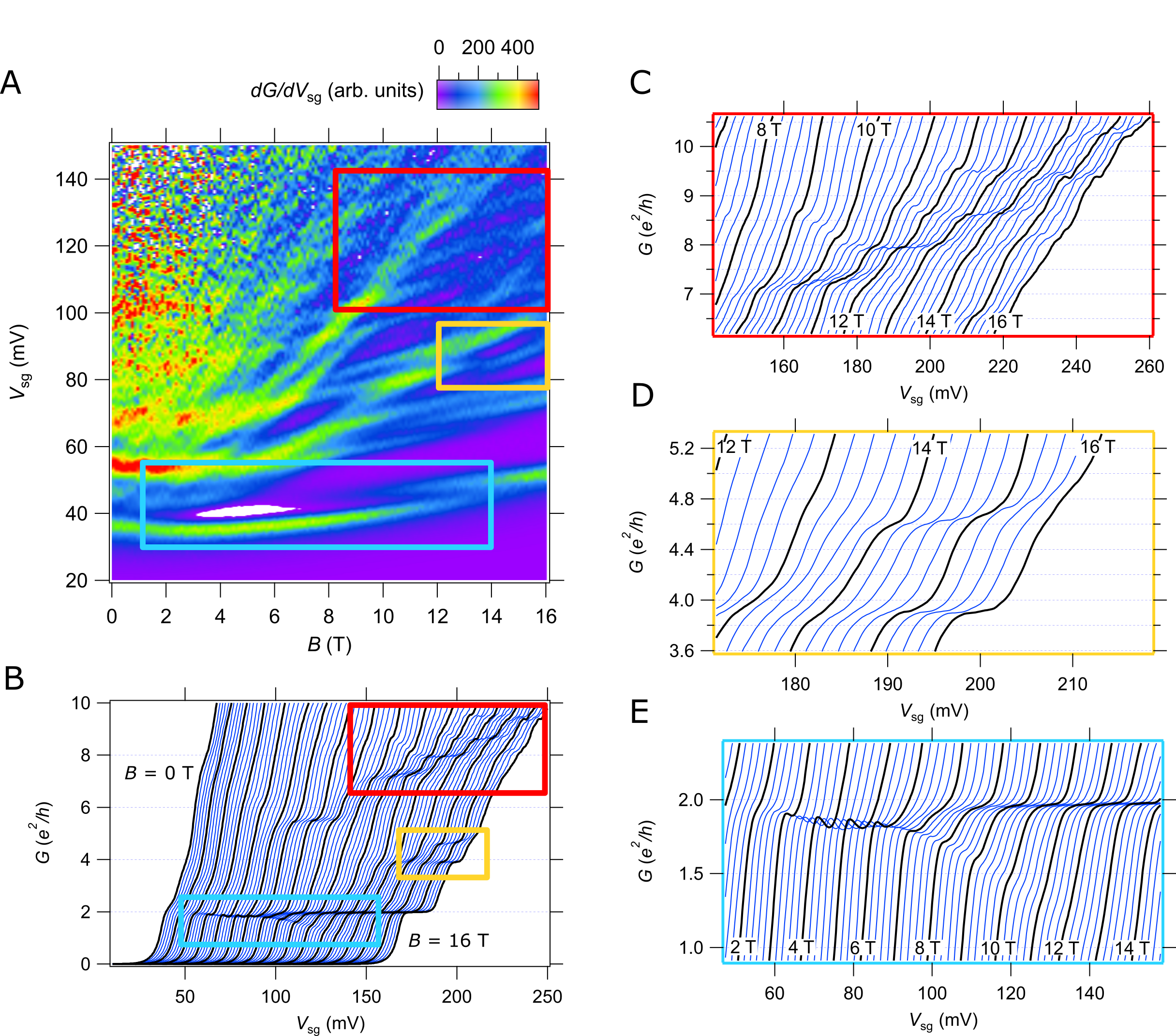}
    \caption{\textbf{Fractional conductance features for vertical superlattice Device A.} \textbf{(A)} Transconductance map with regions highlighted in colored boxes as a guide to the eye to indicate the location of the conductance curves.  \textbf{(E)} Full conductance curves with colored boxes indicating corresponding locations in the transconductance map and other conductance panels. \textbf{(C)} and \textbf{(D)} Conductance curves showing conductance plateaus which correspond to the ``fractured'' states in the transconductance map.  The conductance jump between the plateaus are fractions of the conductance quanta $e^2/h$.  \textbf{(E)} Conductance curves highlighting the fractional conductance feature occurring below the $2~e^2/h$ plateau.}
    \label{fig:fractional_linecuts_A}
\end{figure}

In addition to the plateau at $2~e^2/h$,  many other subband features are readily seen at higher conductance values.  Some of the subbands that make up this additional manifold of states are shown in more detail in Figure \ref{fig:fractional_linecuts_A}.  Conductance curves (Figure \ref{fig:fractional_linecuts_A}C-E) at several parameter values. Corresponding colored boxed, as a guide to the eye, indicate where they exist within the transconductance map (Figure \ref{fig:fractional_linecuts_A}A) and the full range of conductance curves (Figure \ref{fig:fractional_linecuts_A}B). In Figures \ref{fig:fractional_linecuts_A}B and C there are several conductance plateaus visible.  The conductance increases between these plateaus correspond to new subbands, the so-called ``fractured'' states, becoming available in the transconductance map.  These appear to be fractional subbands as the increase in conductance between the plateaus are fractions of the conductance quanta $e^2/h$.  Figure \ref{fig:fractional_linecuts_A}D shows the fractional conductance feature occuring below the $2~e^2/h$ plateau in more detail.  The feature first appears in the form of a conductance peak then bifurcates forming the $\sim 2~e^2/h$ plateau, and a lower fractional conductance feature that evolves with magnetic field.

Data for a second superlattice device, shown in (Figure \ref{fig:deviceB}), yields qualitatively similar behavior.  The overall subband structure resembles the subband structure of an electron waveguide device with no superlattice patterning, but the subbands are, like with Device A, ``fractured'' into additional manifolds with fractional conductance plateaus. Device B also shares the prominent highly quantized conductance plateau at 2 $e^2/h$.

\begin{figure}
    \centering
    \includegraphics[width=\columnwidth]{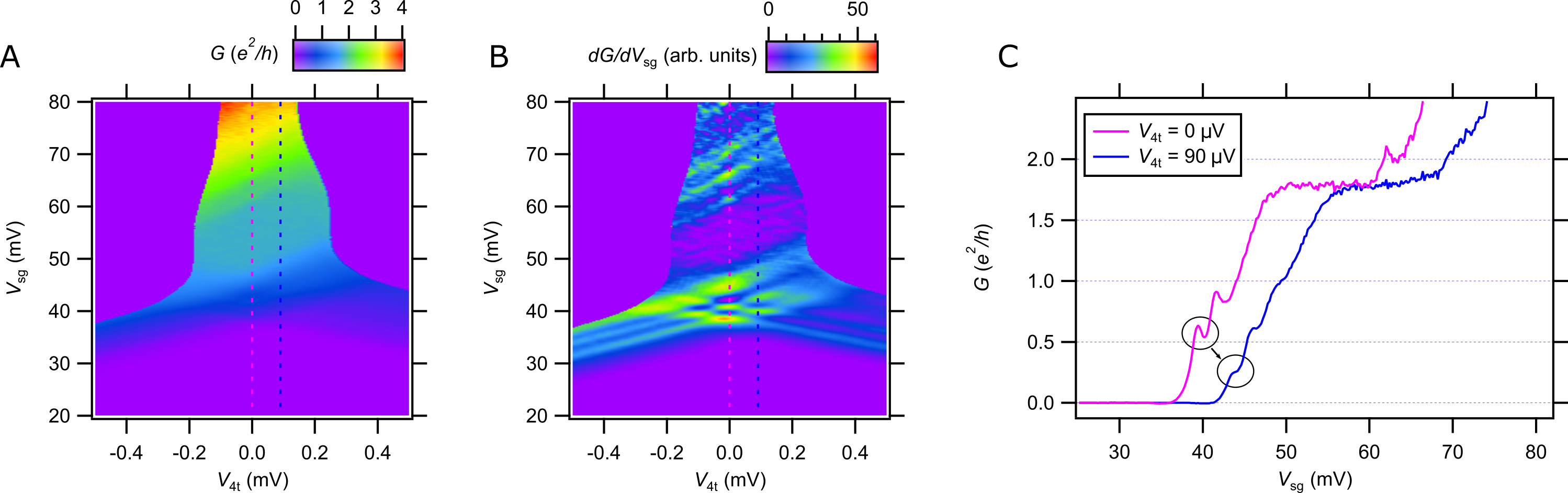}
    \caption{\textbf{Finite-bias spectroscopy for Device A.} \textbf{(A)} Conductance ($G$) intensity map as a function of four-terminal voltage $V_{\mathrm{4t}}$ and side gate voltage $V_{\mathrm{sg}}$, pink and blue dashed lines indicate the locations for the vertical linecuts shown in \textbf{(C)}. \textbf{(B)} Transconductance ($dG/dV_{\mathrm{sg}}$) intensity map as a function of four-terminal voltage $V_{\mathrm{4t}}$ and side gate voltage $V_{\mathrm{sg}}$. The transconductance map shows the diamond features indicating ballistic transport in the superlattice devices.  \textbf{(C)} Vertical conductance linecuts at $V_{\mathrm{4t}}=0$ and $90~\mathrm{\mu V}$.  Circles indicate fractional conductance values below the $\sim2~e^2/h$ plateau (corresponding to the lowest diamond features visible in the transconductance map in panel \textbf{(B)}) that become half of their value at a finite bias. Curves are offset for clarity. Data taken at $B$ = 13 T.}
    \label{fig:DeviceA_IV}
\end{figure}

Finite-bias spectroscopy for 1D superlattice device A (Figure \ref{fig:DeviceA_IV}) reveals a characteristic diamond structure in the transconductance.  This feature is associated with clean ballistic transport \citep{Glazman1989,Patel1990} and is due to unevenly populated subbands at large finite biases which give rise to half-plateaus.  The diamond visible in the transconductance corresponds to a fractional conductance feature below the $2~e^2/h$ plateau at around $0.5~e^2/h$ at zero-bias and about half that value at finite-bias.  The presence of this characteristic diamond structure rules out the likelihood that the fractional features are due to reduced transmission in the channel.

Control devices, straight unmodulated waveguides, discussed in more detail elsewhere \citep{Annadi2018,Briggeman2019}), do not show fractional conductances.  Although not shown here, the behavior of such ``control'' devices consists mainly of conductance plateaus that are quantized in integer values of $e^2/h$, i.e., lacking in the fractionalized subbands observed here.

\section*{Discussion}

Fractional conductances in 1D transport have been reported in a variety of systems, and the phenomenon generally arises when there are strong electron-electron interactions
The fractional quantum Hall state \cite{Tsui1982} is perhaps the best known and investigated example.  The "0.7 anomaly" \citep{Thomas1996} in quantum point contacts \cite{Wees1988,Wharam1988} has been extensively investigated {REF}.  While electron-electron interactions are believed to play a central role in the formation of the conductance plateau observed at $0.7\times(2e^2/h)$ \citep{Bauer2013}.  Fractional conductances have been observed in several 1D quantum wire systems such as strained Ge-based hole quantum wires \citep{Gul2018} and GaAs-based quantum wires \citep{Kumar2019}.  

Shavit et al have considered a theoretical framework\citep{Oreg2014,Shavit2019} in which fractional conductances arise in multi-channel 1D quantum wires due to high order scattering processes.  These processes necessarily require strong repulsive electron-electron interactions. The electron-electron interactions enable momentum-conserving back-scattering processes in the nanowires which lead to fractional conductance states.
This theory predicts a robust plateau at $1.8 e^2/h$, which is due to a coordinated scattering process that explicitly breaks time reversal symmetry.  Our observation of a stable conductance plateau near $1.8 e^2/h$ near $B=2$ T is consistent with this prediction.

A defining characteristic of the \LAOSTO system is the prevalence of strong \textit{attractive} electron-electron interactions \citep{Cheng2015,Cheng2016,Annadi2018,Briggeman2019}.  Both vertical superlattice devices show signs of (weak) superconductivity at $B=0$ T.  Empirically, unperturbed electron waveguides (which possess attractive interactions)  do not exhibit fractional conductance plateaus.
Devices at the \LAOSTO interface exhibit electron pairing without superconductivity \citep{Cheng2015,Annadi2018}.  In electron waveguides, this interaction causes electron subband energy minima to lock together, either near zero magnetic field or at re-entrant values, resulting in conductance steps of $2~e^2/h$.  The superlattice modulation of the electron waveguides is empirically linked to enhanced electron pairing fields.  The effect is significant: superlattice devices have pairing fields of $B_{\mathrm{P}}>16$ T.  Control devices written in the same area of the sample generally show smaller pairing fields $B_{\mathrm{P}}\sim 10$ T \citep{Annadi2018}.  The enhanced pairing strength appears to be a consequence of the potential modulation, although the physical mechanism is unclear.  Superlattices formed by lateral modulation do not show an enhanced pairing field \citep{Briggeman2020L}.

Another effect that is correlated with the vertical modulation is a spin-orbit like effect in the device.  The lowest subband in device A (seen in the transconductance map in Figure \ref{fig:fig2}B) bends upward at zero magnetic field, so that the minima of the lowest subband are at a finite magnetic field.  This may be due to the engineering of a spin-orbit field, and is not usually observed in quasi-1D electron waveguide devices at the \LAOSTO interface.  As the electrons travel through the device with momentum $\vec{k}=k \hat{x}$ the periodic vertical modulation will create an effective electric field $\vec{E}_{\mathrm{eff}}(x)= E_0 \mathrm{sin} (k x) \hat{z}$ which will yield an effective spin-orbit field $\vec{B}_{\mathrm{SO}}\propto \vec{k}\times \vec{E}_{\mathrm{eff}}$ in the $\hat{y}$ direction.  
This could result in a coupling between the spin-up and spin-down particles which may be the mechanism for enhancing the pairing field in these devices.

We present below a minimal model for the pairing of the two first orbitals $|0,0,\downarrow\rangle$ and $|0,0,\uparrow\rangle$ that supports this interpretation, as it predicts an enhanced pairing of these orbitals in the presence of a spin-orbit coupling.
 Our theory is based on a 1D model with an additional simple spin-orbit coupling term of the form $H_\mathrm{SO} = \alpha_\mathrm{SO}k\sigma_y$, with no spatial dependence of the spin-orbit coupling strength $\alpha_\mathrm{SO}$ to simplify the calculations. In the mean-field approximation, our Hamiltonian in momentum space reads
\begin{equation}\label{Hk}
H= \sum_k H_k = \sum_k \bigg[(\xi_{\uparrow k} + \Sigma_\uparrow )c_{\uparrow k}^\dagger c_{\uparrow k} + (\xi_{\downarrow k} + \Sigma_\downarrow )c_{\downarrow k}^\dagger c_{\downarrow k} + \Delta(c_{\uparrow k}^\dagger c_{\downarrow -k}^\dagger - c_{\uparrow k}c_{\downarrow -k}) + i \alpha_\mathrm{SO} k (c_{\uparrow k}^\dagger c_{\downarrow k} - c_{\downarrow k}^\dagger c_{\uparrow k})\bigg],
\end{equation}
where $\xi_{\sigma k}$ ($\sigma = \downarrow, \uparrow$) are the single-particle energies of the states $|0,0,\downarrow\rangle\otimes|k\rangle$ and $|0,0,\uparrow\rangle\otimes|k\rangle$, and where $\Sigma_\sigma$ and $\Delta$ are respectively the Hartree shifts and the pairing order parameter defined as
\begin{align}
&\Sigma_\sigma = \frac{U}{2 \pi} \int \langle c_{\bar{\sigma}k}^\dagger c_{\bar{\sigma} k}\rangle \mathrm{d}k, \quad \quad \Delta=\frac{U}{2 \pi} \int \langle c_{\downarrow-k} c_{\uparrow k}\rangle  \mathrm{d}k,
\end{align}
where $U$ is the electron-electron interaction. The mean-fields are found self-consistently, and indicate the presence of electrons and paired electrons in the waveguide. Calculating $\Delta$ and other single-particle correlation functions (see supplemental information) determines phase diagrams for different values of $\alpha_\mathrm{SO}$ and $U$. Our results are shown in  Figure~\ref{fig:FigMF}. 
An enhanced pairing area, defined as the region of non-zero $\Delta$, is obtained for increasing $\alpha_\mathrm{SO}$. This minimal model (which could be extended to position-dependent potentials and spin-orbit coupling) thus supports the idea that a spin-orbit coupling engineered by the experimental setup increases the pairing of the two lowest subbands $|0,0,\downarrow\rangle$ and $|0,0,\uparrow\rangle$ into a singlet state, yielding a first conductance step of $2e^2/h$. 
Future experiments applying an in-plane magnetic field could test this theory. A similar effect is seem in lateral 1D superlattice devices \citep{Briggeman2020L}.

The ability to create new superlattice structures, and modulate interactions in 1D systems, opens new frontiers in the development of quantum matter.  The systems created here focus on low-dimensional confined structures, which are challenging to create using other methods.  The regular superlattice structure can be replaced with quastiperiodic order, artificially imposed disorder, topological defects, or combined with lateral perturbations, to name just a few possibilities.  Unlike the Kronig-Penney description, electron-electron interactions play a defining role in the resulting quantum phases, and future discoveries of emergent phases in this family of 1D systems are highly likely.

\section*{Methods}

3.4 unit cell (u.c.) \LAOSTO samples were grown using pulsed laser deposition (PLD) (described in more detail elsewhere \citep{Cheng2011}).  Electrical contact was made to the interface by ion milling  through the interface and backfilling with Ti/Au.  C-AFM writing was performed by applying a voltage bias between the AFM tip and the interface, with a 1 G$\mathrm{\Omega}$ resistor in series.  Writing was performed in 30-40\% relative humidity using an Asylum MFP3D AFM.  Written samples were then transferred into a dilution refrigerator and cooled to a base temperature of $\sim25$ mK.  4-terminal measurements were performed using standard lockin techniques at a frequency of 11 Hz with an oscillation amplitude of 1 mV. 4-terminal I-V curves were performed by applying a DC source-drain bias across the device.

\bibliographystyle{apsrev4-1}

\section*{Acknowledgements}

JL acknowledges support from a Vannevar Bush Faculty Fellowship, (N00014-15-1-2847) and the National Science Foundation (PHY-1913034). Work at the University of Wisconsin was supported by funding from the DOE Office of Basic Energy Sciences under award number DE-FG02-06ER46327 (C.B.E). F. D. and A. J. D. acknowledge support from the EPSRC Programme Grant DesOEQ (EP/P009565/1) and the AFOSR (FA9550-18-1-0064).

\section*{Author contributions statement}

M.B., P.I., and J. Levy conducted the experiments. F.D., E.M., and A.J.D. made the theoretical model calculations. J.Li and M.H. processed the samples. H.L., J.-W.L., K.E. and C.-B.E. synthesized the thin films and performed structural and electrical characterizations. All authors reviewed the manuscript.

\section*{Additional information}


\textbf{Competing interests} The authors declare no competing interests. 


\begin{figure}
    \centering
    \includegraphics[width=\columnwidth]{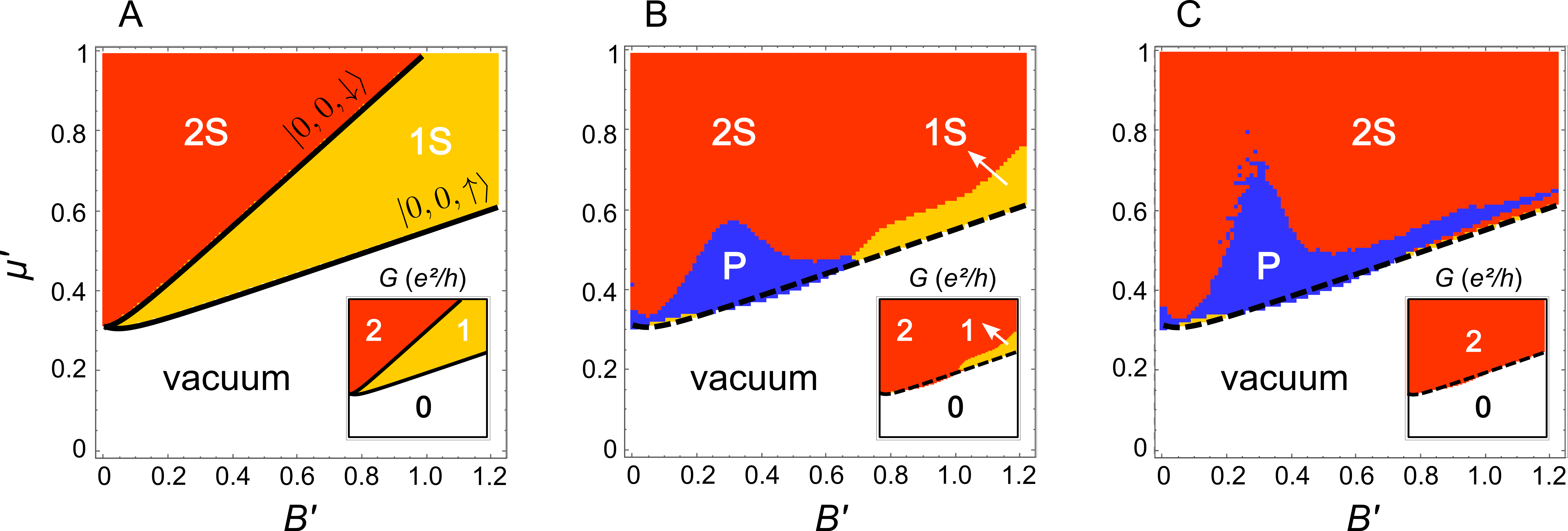}
    \caption{\textbf{Phase diagrams of the mean-field model.} The diagrams A, B and C show the different phases of the waveguide obtained from the mean-field model~(\ref{Hk}) as a function of the dimensionless magnetic field $B'$ and chemical potential $\mu'$. \textbf{(A)}: No interaction ($U = 0$) and no spin-orbit coupling ($\alpha_\mathrm{SO} = 0$). The areas labelled as "vacuum", "1S" and "2S" correspond respectively to an empty phase, a single-particle phase and a two-single-particle phase. They are delimited by the Zeeman-splitted single-particle energies $\xi_{\sigma k}(B', \mu')$ of the states $|0,0,\downarrow\rangle\otimes|k\rangle$ and $|0,0,\uparrow\rangle\otimes|k\rangle$  (see Supplemental Information for detailed expressions) satisfying $\xi_{\sigma 0}(B', \mu') = 0$ (solid black lines). The inset shows the associated conductance. \textbf{(B)}: Finite dimensionless interaction ($U' = 0.2$) and no spin-orbit coupling ($\alpha_\mathrm{SO} = 0$). The presence of interactions induces a pairing $\Delta$ of the orbitals in some region of parameters (phase "P"), moving the splitting of the two single-particle bands to higher magnetic field ($B' \sim 0.7$). The dashed black line corresponding to $\xi_{\uparrow 0}(B', \mu') = 0$ is left there as a guide for the eye. \textbf{(C)}: Finite dimensionless interaction ($U' = 0.2$) and spin-orbit coupling ($\alpha'_\mathrm{SO} = 0.12$). The presence of the spin-orbit coupling increases the area of the paired phase up to $B' \sim 1.2$. This phenomenon is not sensitive to the specific parameter values.}
    \label{fig:FigMF}
\end{figure}

\beginsupplement

\section{Supplemental information}

\subsection{Device writing and measurement parameters}

All three devices discussed were written in the same location on the \LAO surface during different cooldowns.  1D vertical superlattice device A and B were written with the same c-AFM writing parameters.  The main channel of the device was written with a tip voltage $V_{\mathrm{tip}}=12$ V at a speed of 50 nm/s.  The barriers for the electron waveguide were $L_{\mathrm{B}}=5$ nm separated by $L_{\mathrm{S}}=10$ nm and created by applying negative voltage pulses of -9 V at a speed of 5 nm/s.  The superlattice was created by applying a tip voltage $V_{\mathrm{tip}}(x)=5 \mathrm{V} \sin((\pi/5~\mathrm{nm}) x)$ with 18 periods.  The control device, Device C, consisted of only the electron waveguide section written with the same parameters.

Data was taken at base temperature of a dilution refrigerator $T\sim30$ mK.  Transport data for 1D superlattice devices was taken using standard lockin techniques with an oscillation amplitude of 1 mV ($250~\mathrm{\mu V}$) at a reference frequency of 11 (13.46) Hz for Devices A (B).  Data displayed for Device C is the zero bias cut of a finite bias spectroscopy.  More examples of typical electron waveguide devices can be found in ref \citep{Annadi2018,Briggeman2019}.

\subsection{Device B}

The transport in Device B did not appear to be as clean as Device A.  There was also an issue with the side gate leaking at low side-gate and magnetic field values, causing the data to be distorted.

\subsection{Mean-field model}

The single-particle energies $\xi_{\sigma k}$ of the states $|m,n,\sigma \rangle \otimes |k\rangle$ are given by~\cite{Annadi2018}
\begin{equation}
\xi_{\sigma k} = \frac{\hbar^2 k^2}{2m_x^*}\left(1 - \frac{\omega_c^2}{\Omega^2}\right) - g \mu_B B s +  \hbar \Omega \left(m + \frac{1}{2}\right) + \hbar \omega_z\left((2n+1) + \frac{1}{2}\right) -\mu ,
\end{equation}
where $\omega_c =eB/m_y^*$ is the cyclotron frequency, $m_x^*$ ($m_y^*$) the effective mass of the electron along the $x$ ($y$)-direction, $\omega_z$ ($\Omega = \sqrt{\omega_y^2 + \omega_c^2}$) the trapping frequency along the vertical (lateral) direction, $g$ the Land\'e factor, $\mu_B$ the Bohr magneton and $s = -1/2$ ($1/2$) for $\sigma = \downarrow$ ($\uparrow$). In Figure~\ref{fig:FigMF}, we defined the dimensionless parameters $\mu' = \mu/E_0$, $B' = B/[E_0 m_y^*/(e \hbar)]$, $U' = U/E_0$ and $\alpha'_\mathrm{SO} = \alpha_{\mathrm{SO}} / (E_0 a)$, in terms of energy $E_0 = \hbar^2/(\pi m_x^* a^2)$ and lattice spacing $a$, and used $m=n=0$, $g = 0.6$, $\hbar \omega_z = 0.18 E_0$, $\hbar \omega_y = 0.07 E_0$ and $m_y^* = 1.5 m_e$ where $m_e$ is the electron mass.

The eigensystem~(\ref{Hk}) was solved iteratively in the electron-hole basis $(c_{\uparrow k}, c_{\uparrow -k}^\dagger,c_{\downarrow k}, c_{\downarrow -k}^\dagger)$ until convergence of the mean-fields. Single-particle phases were identified from non-zero values of $\int \langle d_{\sigma k}^\dagger d_{\sigma k} \rangle  \mathrm{d}k$ (taken in thermal equilibrium at zero temperature) where $d_{\pm k}$ are the single-particle annihilation operators that diagonalize the non-interacting version of~(\ref{Hk}) as $H = \sum_{k,\pm} E_{\pm k} d_{\pm k}^{\dagger} d_{\pm k}$ with $E_{\pm k} = (\xi_{\downarrow k}+ \xi_{\uparrow k} \pm \sqrt{4 k^2 \alpha_\mathrm{SO}^2 + (\xi_{\downarrow k} - \xi_{\uparrow k})^2})/2$. 
The pair phase in Figure~\ref{fig:FigMF} corresponds to the regime where $\Delta > 10^{-2}E_0$.

\begin{figure}[h]
    \centering
    \includegraphics[width=\columnwidth]{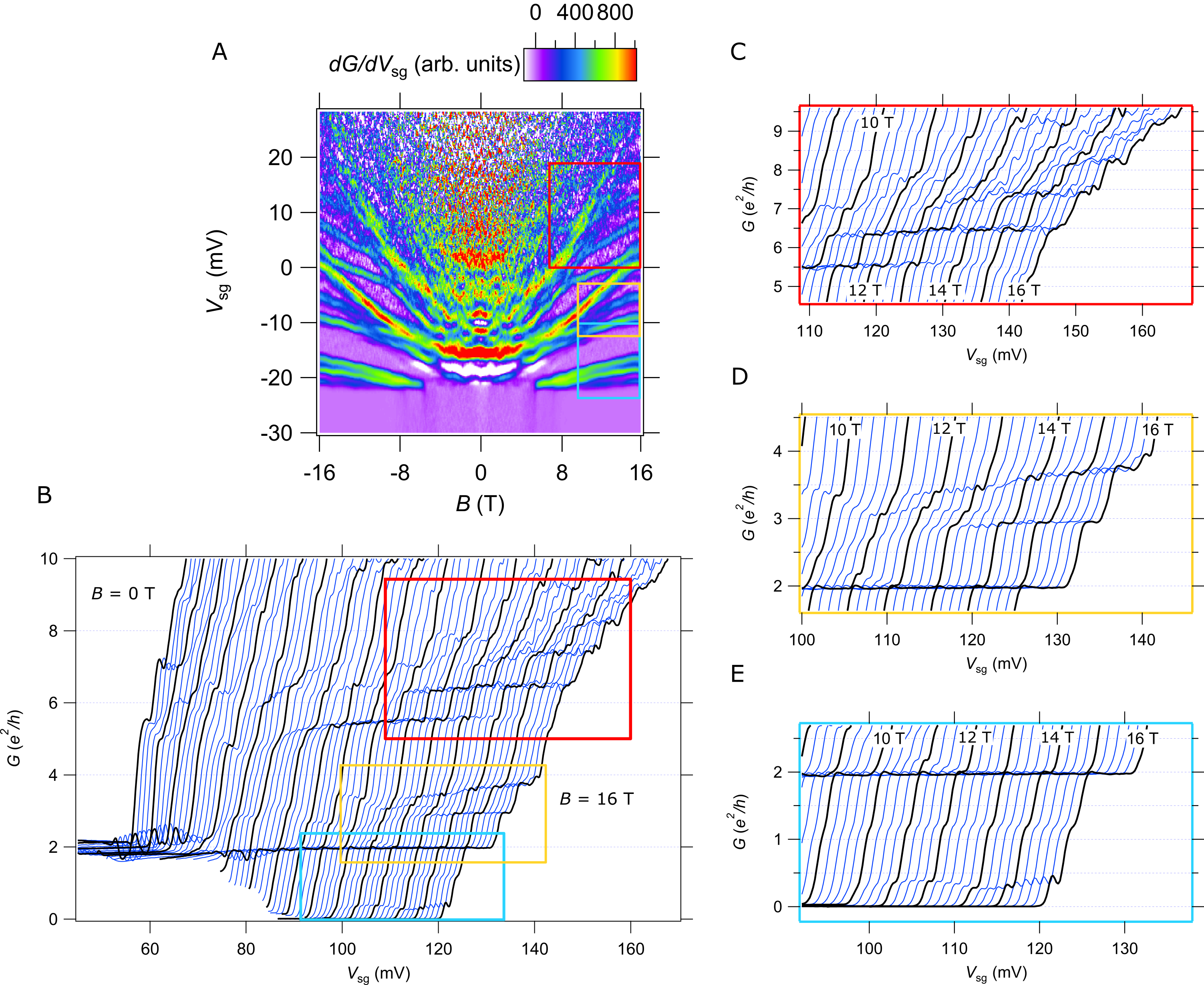}
    \caption{\textbf{Magnetotransport data for vertical superlattice device B.}  \textbf{(A)} Transconductance map $dG/dV_{\mathrm{sg}}$ as a function of side gate voltage $V_{\mathrm{sg}}$ and magnetic field $B$.  Purple regions indicate conductance plateaus, zero transconductance.  Red/yellow/green/blue regions indicate increases in conductance when new subbands become available. White regions indicate negative transconductance. Colored boxes are guides to the eye indication the location of highlighted conductance curves.  \textbf{(B)} Plot showing full conductance data.  Conductance curves at 1T intervals are highlighted in black and are offset clarity.  \textbf{(C)}-\textbf{(E)} Conductance $G$ as a function of side gate voltage $V_{\mathrm{sg}}$ curves at different out-of-plane magnetic field $B$ values highlighting some fractional conductance features.}
    \label{fig:deviceB}
\end{figure}

\end{document}